\documentclass[10pt,twocolumn,twoside]{elsarticle}

\usepackage{lineno,hyperref}
\usepackage{xcolor}
\modulolinenumbers[5]

\usepackage{amsmath,amssymb,amsthm,mathtools}
\usepackage{multicol}
\usepackage{cuted}      
\usepackage{microtype}  

\usepackage{graphicx}
\usepackage{caption}
\usepackage{subcaption}
\usepackage{float}

\usepackage{geometry}
\usepackage{times}

\def\DOI{DOI:10.1016/j.optlaseng.2026.109664}%
\def\AcceptedDate{25 Jan. 2026}

\makeatletter
\def\ps@pprintTitle{%
     \let\@oddhead\@empty
     \let\@evenhead\@empty
     \def\@oddfoot
       {\hbox to \textwidth%
        {\ifnopreprintline\relax\else
        \@myfooterfont%
         \ifx\@elsarticlemyfooteralign\@elsarticlemyfooteraligncenter%
           \hfil\@elsarticlemyfooter\hfil%
         \else%
         \ifx\@elsarticlemyfooteralign\@elsarticlemyfooteralignleft%
           \@elsarticlemyfooter\hfill{}%
         \else%
         \ifx\@elsarticlemyfooteralign\@elsarticlemyfooteralignright%
           {}\hfill\@elsarticlemyfooter%
         \else%
           {\rm
           \parbox[b]{0.85\textwidth}{%
             Accepted in \@journal. \copyright 2026 Elsevier (CC BY-NC-ND 4.0). \DOI
           }\hfil \AcceptedDate}\fi%
         \fi%
         \fi%
         \fi%
         }
       }%
     \let\@evenfoot\@oddfoot}
\makeatother
\journal{Optics and Lasers in Engineering}

\newcommand{\EQ}[1]{Eq.~(\ref{#1})}
\newcommand{\EQtwo}[2]{Eqs.~(\ref{#1}) and (\ref{#2})}
\newcommand{\EQthree}[3]{Eqs.~(\ref{#1}), (\ref{#2}) and (\ref{#3})}

\renewcommand{\Vec}[1]{{\boldsymbol{#1}}}

\newcommand{\Vr}{\Vec{r}}
\newcommand{\Vs}{\Vec{s}}%
\newcommand{\Vk}{\Vec{k}}%
\newcommand{\Vn}{\Vec{n}}%

\newcommand{\Vkbg}{\Vk_{\rm bg}}%

\newcommand{\WrapSymbol}{\mbox{${\mathcal W}$}}
\newcommand{\UnwrapSymbol}{\mbox{${\mathcal U}$}}
\newcommand{\Wrap}[1]{\WrapSymbol\left\{#1\right\}}
\newcommand{\Unwrap}[1]{\UnwrapSymbol\left\{#1\right\}}

\newcommand{\inN}{\in{\mathbb{N}}}

\newcommand{\Fourier}{{\mathcal F}}
\newcommand{\FT}[1]{\Fourier\!\{#1\}}

\newcommand{\FTinv}[1]{\Fourier^{-1}\!\{#1\}}

\newcommand{\Ave}[1]{\left\langle #1\right\rangle}
\newcommand{\RMS}[1]{{\rm RMS}\{#1\}}

\def\ds{\epsilon}
\def\dss#1{\epsilon_{#1}}
\def\dsamp{\delta_{x}}
\def\fM{f^{\rm M}}
\def\IM{I^{\rm M}}
\def\fM{f^{\rm M}}
\def\IM{I^{\rm M}}

\def\sI{S}
\def\sIn#1{S_{#1}}

\def\sIv#1{{\boldsymbol{S}}^{#1}}
\def\Nat#1{{\rm Rec}\left\{#1\right\}}
\def\Recx#1{{\rm Rec}_x\left\{#1\right\}}
\def\Recy#1{{\rm Rec}_y\left\{#1\right\}}
\def\dsn#1{\ds_{#1}}
\def\hatds{\hat{\ds}}
\def\hatA{\hat{A}}
\def\hatdsn#1{\hat{\ds}_{#1}}
\def\hatphi{\hat{\phi}}
\def\hatd{\hat{d}}

\def\Ave#1{\left\langle #1\right\rangle}
\def\RMS#1{{\rm RMS}\{#1\}}
\def\Norm#1{\left|\!\left|#1\right|\!\right|}
\def\Erel#1{E_{\rm rel}\left\{#1\right\}}
\def\Ereltheo#1{\tilde{E}_{\rm rel}\left\{#1\right\}}

\newlength\HsizeInTwocol
\newif\ifMODIFY\MODIFYfalse

\usepackage{xcolor} 
\usepackage{cancel}

\ifMODIFY
  \definecolor{darkgreen}{rgb}{0.,0.6,0}
  \definecolor{Blue}{rgb}{0,0,1}
  \usepackage{ulem}%
  \usepackage{cancel}%
  \newcommand{\MODIFIED}[2]{{\color{red}{#1}}{%
    \setbox0\hbox{#2}\ifdim\wd0>0pt\color{blue}{\sout{#2}}\fi
    }%
  }%
  \newcommand{\Modified}[2]{{\color{red}{#1}}{%
    \setbox0\hbox{#2}\ifdim\wd0>0pt\color{blue}{\,\xcancel{#2}}\fi
    }%
  }%
  \newcommand{\MathModified}[2]{\Modified{\mbox{$#1$}}{\mbox{$#2$}}}
  \newcommand{\MEMO}[1]{{\color{cyan}{\bf #1}}}%
  \newcommand{\SamiaResponse}[1]{{\color{darkgreen}{\bf #1}}}%
\else
  \newcommand{\MODIFIED}[2]{{\color{black}{#1}}}%
  \newcommand{\Modified}[2]{{\color{black}{#1}}}%
  \newcommand{\MathModified}[2]{\Modified{\mbox{$#1$}}{\mbox{$#2$}}}
  \newcommand{\MEMO}[1]{}%
  \newcommand{\SamiaResponse}[1]1{}%
\fi

\begin{document}



\onecolumn
\begin{strip}

\begin{frontmatter}

\title{Wavefront Reconstruction for Fractional Lateral Shear Measurements using Weighted Integer Shear Averages}

\author[myfirstaddress]{Samia Heshmat}
\author[mysecondaryaddress]{Satoshi Tomioka \corref{mycorrespondingauthor}}
\cortext[mycorrespondingauthor]{Corresponding author}
\ead{tom@qe.eng.hokudai.ac.jp}
\author[mysecondaryaddress]{Naoki Miyamoto}
\author[mysecondaryaddress]{Yuji Yamauchi}
\author[mysecondaryaddress]{Yutaka Matsumoto}
\author[mysecondaryaddress]{Naoki Higashi}
\address[myfirstaddress]{Faculty of Engineering, Aswan University, Aswan, 81542, Egypt}
\address[mysecondaryaddress]{Faculty of Engineering, Hokkaido University, Sapporo, 060-8628, Japan}

\begin{abstract}
Wavefront reconstruction in lateral shearing interferometry
typically 
assumes that the shear amount is an integer
multiple of the sampling interval.
When the shear is fractional, approximating it with
the nearest integer value leads to noticeable
reconstruction errors. To address this,
we propose a weighted integer shear averaging method.
The approach combines reconstructions from nearby integer
shears with carefully chosen weights designed to cancel the dominant
error terms. Analytical error analysis shows
that two-shear averaging removes first-order errors, while
three-shear averaging removes second-order errors.
Numerical simulations with a test wavefront confirm
that the method achieves significantly lower RMS error
than conventional single-shear reconstruction.
The technique is simple, computationally efficient,
and can be readily extended to two-dimensional interferometry.
This makes weighted integer shear averaging a practical
and accurate tool for wavefront reconstruction when fractional shear is unavoidable.

\end{abstract}


\end{frontmatter}

\end{strip}
\twocolumn

\section{Introduction}
\label{sec:intro}

Interferometers are widely employed for wavefront measurement and 
optical metrology applications~\cite{Yoshizawa2015, Sirohi2009, Schnars2015}.
In conventional two-beam interferometers, such as the Michelson or 
Mach-Zehnder configuration, the object beam and the reference beam travel 
along separate optical paths.
This path separation makes such a system highly 
sensitive to environmental disturbances such as vibrations, temperature 
gradients and air turbulence, which are difficult to correct 
in practice \cite{Mana:18, Zhang:11}. 
Moreover, because the beam must be split and recombined, 
numerous optical components are required, which further increases 
system complexity and susceptibility to noise sources. 

For example, in certain measurement systems, 
such as a computed tomography (CT) or a tomographic phase microscopy, 
wavefront measurements are performed by changing the angle of incidence 
of the probing beam \cite{Bon:09, Tomioka:15, Tomioka:17, Ling:17}. In such cases, vibrations 
and misalignments from each optical element act independently, 
further complicating reliable interference measurements.
In contrast, a shear interferometer offers a more robust approach:
the object wavefront after transmission or reflection from 
the object under test is split into two parallel beams: 
the original object beam and a laterally displaced replica (the shear beam)
\cite{Murty:64, Riley:77}. 
The two beams interfere with each other, 
sharing almost the same optical paths and requiring 
fewer optical elements than conventional two-beam interferometers, 
thereby mitigating environmental sensitivity and system complexity. 

However, unlike the two-beam interferometers, the fringes recorded in a shear 
interferometer encodes only the wavefront difference
between 
the two beams, rather than the absolute wavefront itself. 
Consequently,  numerical processing is 
required to reconstruct the actual object wavefront 
from this differential measurement. 
Mathematically, the relationship between the measured differential phase 
$\fM(x,s)$ for a lateral shear $s$ and 
the original object wavefront $\phi(x)$ can be expressed as
\begin{align}
  \label{eq:difference}
  \fM(x,s) = \phi(x+s) - \phi(x).
\end{align}

The methods developed to address the wavefront reconstruction problem 
in lateral shearing interferometry, it can be broadly classified into modal and
zonal approaches.
The latter is further classified into two groups:
approaches based on the least-squares method in the spatial domain, 
and approaches based on the Fourier transform (FT).
In modal methods, the unknown wavefront $\phi(x)$ is expanded using orthogonal basis functions, 
typically Zernike polynomials~\cite{Rimmer:75, Harbers:96, Okuda:00, Dai:13, Mochi:15, Ling:15, Shi:23, Luo:25}. 
These methods are robust against measurement noise but tend to smooth out 
high-frequency or localized wavefront variations. 
In the zonal methods with the least-squares method in 
the spatial domain~\cite{Southwell:80, Zou:00, Dai:12, Dai:16, Zhai:16},
the reconstruction problem is formulated as a system of equations where 
the variables to be solved are wavefronts at the measured pixels.
Solving the equations requires a high computational cost 
because the number of variables is large.
Moreover, these methods suffer from rank-deficient problems.
They can be solved by using multi-shear methods, which employ
two or more measurements of phase difference with different shear amounts.
In contrast,
the zonal methods based on FT~\cite{Hudgin:77, Freischlad:86}
do \MODIFIED{not}{no} need to solve simultaneous equations,
since
they only require
a FT, multiplication by a factor, and an inverse FT,
such as
$\phi(x)=\FTinv{\FT{\fM(x;s)}/(e^{iks}-1)}$
where $\FT{}$ means FT operator
with the basis function $e^{-ikx}$.

However, FT-based reconstruction suffers from several inherent limitations 
regarding reconstruction errors.
The first is the loss of the $1/s$ harmonic components 
in the spectral domain. 
To address this problem, 
the use of the Tikhonov regularization~\cite{Servin:96},
or multi-shears methods~\cite{Guo:12,Guo:14}
similar as the zonal methods based on least squares in the spatial domain,
are proposed.
The second major problem is that
FT is based on
periodic boundaries,
which introduces spectral leakage 
when the underlying phase is aperiodic. 
The natural extension technique proposed by 
Elster~\textit{et al.}~\cite{Elster:99,Elster:99-B,Elster:00}
mitigates boundary artifacts by introducing
mathematically consistent 
domain padding, which has been 
applied to many applications~\cite{Dubra:04,Falldorf:07,Zhai:17,Zhai:17-high,Ling:17}.
Tomioka~\textit{et al.}~\cite{Tomioka:23} proposed a coupling method of 
the spectral interpolation method
to solve the loss of $1/s$ harmonics problem
with the natural extension method
to solve the periodicity problem for one-dimensional reconstruction.
This method requires only a single 
differential phase measurement in one dimension. 
For two-dimensional reconstruction, it also couples 
the least-squares method~\cite{Tian:95} to solve ambiguities of a uniform wavefront, called the piston term. 
However, 
a limitation remains regarding shear amounts.

The basic principle of lateral shearing interferometry and 
most of the wavefront reconstruction methods 
assume that
the shear applied during measurement
is an integer multiple of the spatial sampling interval $\dsamp$.
However, in real-world systems,
it is often impractical to constrain shear values
to discrete grid-aligned steps.
Limitations in mechanical or electronic control,
or the need for higher resolution,
frequently result in the application of non-integer (fractional) shears.
This mismatch introduces a significant challenge.
In this manuscript, 
we provide 
a detailed theoretical analysis of the 
wavefront reconstruction errors,
deriving expressions for the reconstruction error introduced by using a nearby
integer shear instead of the true fractional value.
The error can be represented as a series expansion
in terms of the normalized shear difference,
with leading terms that are proportional to the first and second powers of the deviation.
Furthermore, we build upon that theoretical foundation
to develop a practical method to reduce the reconstruction error
in fractional shear measurements.

This manuscript is structured as follows. 
Section~\ref{sec:theory} reviews 
the shear interference fringe analysis process.
Section~\ref{sec:algorithm} summarizes the analytical error expressions for 
fractional shear substitution. In addition, this section
introduces the proposed multiple integer shear averaging approach and 
provides the mathematical basis for its effectiveness. 
Section ~\ref{sec:results} presents numerical simulations 
validating the proposed approach, 
and Section~\ref{sec:conclusion} concludes with a discussion of future directions.

\section{Shear interference fringe analysis process}
\label{sec:theory}

Unlike a conventional two-beam interferometer that compares 
an object wavefront with a separate reference beam, 
a shear interferometer laterally displaces the object beam itself 
to generate interference. 
Consequently, the recorded fringes represent the phase difference 
between two laterally shifted replicas of the same wavefront  
rather than the absolute phase distribution. 
This makes direct interpretation of the measured intensity pattern 
nontrivial and necessitates numerical reconstruction to recover 
the original wavefront.
The measured fringe pattern $\IM(\Vr;\Vs)$ in the lateral shear interferometry with a shear amount $\Vs$
can be simply  expressed as 
\begin{align}
\label{eq:fringe1}
  \IM(\Vr;\Vs) &= I_0(\Vr)\left[1+\cos\!\left(\Vkbg\!\cdot\!\Vr+\delta\phi(\Vr;\Vs)\right)\right],
\end{align}
where $\Vkbg$ denotes the spatial carrier 
frequency determined by
the difference between the wavenumber vectors in vacuum of the two beams as 
$\Vkbg=\Vk_0(\Vr+\Vs)-\Vk_0(\Vr)$ which is almost uniform,
and $\delta\phi(\Vr;\Vs)$ represents the phase difference between the two beams
represented as
\begin{align}
  \label{eq:fringe_shear}
  \delta\phi(\Vr;\Vs)=\phi(\Vr+\Vs)-\phi(\Vr).
\end{align}

Thus, shear interferometry provides phase differences that require 
specialized reconstruction algorithms, which are the main focus of this study.
\EQ{eq:fringe1} means that the parallel 
stripes are distorted by non-uniform $\delta\phi(\Vr;\Vs)$.
Once $\IM(\Vr;\Vs)$ is obtained through measurement,
$e^{j\delta\phi(\Vr;\Vs)}$ can be extracted in the same way
as a two-beam interferometer by the FT method \cite{Takeda:82}.
After that, the phase difference can be obtained by taking a complex argument.
However, the obtained phase is a wrapped phase including $2\pi$ phase jumps.
\begin{align}
\Wrap{\delta\phi(\Vr;\Vs)} = \arg\left\{ I_0(\Vr) e^{j\delta\phi(\MathModified{\scriptsize\Vr}{\scriptsize\Vr\;\Vs})}\right\},
\end{align}
where $\Wrap{}$ shows the wrapping operator.
To solve the phase jump problem,
a phase unwrapping process, which is the same as the case of the two-beam interferometer, must be applied.
Several unwrapping methods have been proposed;
among them the localized compensator phase unwrapping method \cite{Tomioka:12}
has demonstrated superior performance \cite{Samia:22}.
The unwrapped phase can then be expressed as
\begin{align}
  \fM(\Vr;\Vs) = \Unwrap{\Wrap{\delta\phi(\Vr;\Vs)}},
\end{align}
where $\Unwrap{}$ shows the unwrapping operator,
and the superscript `M' means quantities obtained from measured fringe 
$\IM(\Vr;s)$ through the aforementioned processes.

In practice, two-dimensional (2D) reconstruction is frequently reformulated into 
one-dimensional (1D) differential equations, where the wavefront $\phi(x)$ 
is recovered from the measured shear data $\fM(x;s)$ defined 
in \EQ{eq:difference}.
Although the method proposed in this paper is
also for reconstructing 1D shear problems,
it can be extended to 2D wavefront reconstruction
by a coupling method which can resolve ambiguities called piston terms included in the reconstructed wavefront.
In practical implementations,
phase differences along two orthogonal $x$- and $y$-directions are measured independently.
Specifically, by setting the shear only along the $x$-direction ($\Vs_x=(s_x,0)$),
the shear effect is restricted to the $x$-direction.
The corresponding differential wavefront, $\fM_x(\Vr,\Vs_x)$,
is obtained via the aforementioned FT and phase unwrapping methods.
Consequently, 1D reconstruction can be performed independently for each $y$ position.
However, this line-by-line reconstruction introduces an ambiguity known as the piston term,
$\overline{\phi_{x}}(y)$, which varies with $y$, as
\begin{align}
  \Recx{\fM_x(\Vr,\Vs_x);s_x)}=\phi(\Vr)+\overline{\phi_x}(y),
\end{align}
where $\Recx{}$ denotes an ideal $x$-directional reconstruction operator using shear amount of $s_x$.
Similarly, for the case of $y$-directional shear
\begin{align}
  \Recy{\fM_y(\Vr,\Vs_y);s_y}=\phi(\Vr)+\overline{\phi_y}(x),
\end{align}
To obtain the wavefront $\phi(\Vr)$, two ambiguities of $\overline{\phi_x}(y)$ and $\overline{\phi_y}(x)$ must be determined.
This problem of the ambiguities 
can be solved by coupling with the method proposed by Tian~{\it et~al.}~\cite{Tian:95}.
Thus, we focus on the 1D reconstruction.
In the following, the dependencies of $y$ are omitted as $\Vr\to x$ and $\Vs\to s$ for simplicity.

If
the shear during measurement is an integer shear ($s=\sI$),
the same shear amount $\sI$ can be used,
so it holds:
 \begin{align}
\phi_{\sI;\sI}(x)=\Nat{\fM(x,\sI);\sI}=\phi(x),
\end{align}
where the $\Nat{}$ operator represents the reconstruction operator
to obtain the phase from the measured phase difference,
and the first and second arguments express
the measured phase difference applied to the operator
and the integer shear used for the reconstruction, respectively.
When we employ the spectral interpolation method
\cite{Tomioka:23} for $\Nat{}$, the reconstructed phase
includes errors due to shear harmonics of $\phi(x)$,
but we choose $s$ such that this is negligible,
and we consider the errors to be negligible.
On the other hand, if the shear during measurement is non-integer,
\begin{align}
\phi_{s;\sI}(x)=\Nat{\fM(x,s);\sI}\ne \phi(x).
\end{align}
Since $s$ and $\sI$ differ,
the reconstructed wavefront $\phi_{s;\sI}(x)$
contains a significant error relative to 
the original wavefront $\phi(x)$.

\section{Dominant error reduction using multiple integer-shear averaging}%

\label{sec:algorithm}

To mitigate reconstruction errors arising from
fractional shear measurements,
we introduce a weighted integer shear averaging method.
The core idea is to suppress the dominant error components present in $\phi_{s;\sI}(x)$
by utilizing the properties of integer shear substitutions.
To clarify the numerical reconstruction from fractional-shear measurements, 
we first define $\phi_{s;\sI}^{\dagger}(x)$ 
as the scaled integer-shear approximation of the wavefront, 
obtained by substituting the fractional shear $s$ with a nearby integer shear 
$S$. This reconstructed function preserves the global structure of 
the true wavefront, but generally contains residual error components 
whose amplitude depends on 
the difference between the fractional shear and the integer shear. 
The key observation underlying the proposed method is that 
when several symmetric integer shears with $S_q$, where $q$ 
can be, for example 
$\{1,2\}$, are selected around $s$, their corresponding reconstructions 
$\phi^{\dagger}_{s;\sIn{q}}(x)$ exhibit dominant error components 
with opposite phases. Therefore, by averaging these wavefronts, 
the leading error terms are cancelled while the true wavefront 
contribution is retained. In the subsections that follow, 
we first analyze the reconstruction errors produced when 
a single integer shear is used, and then demonstrate how these errors 
can be reduced by systematically averaging reconstructions 
obtained from several integer shears, $S_q$,
located near the target fractional shear, $s$.

\subsection{Reconstruction error using a single integer shear}
\label{sec:3-1-int-shear-dagger}
In this subsection, we introduce a scaled wavefront $\phi_{s;\sI}^\dagger(x)$ designed to reduce the error inherent in $\phi_{s;\sI}(x)$.
Since the direct error estimation of $\phi_{s;\sI}^\dagger(x)$ is analytically complex,
we also introduce an approximated scaled wavefront $\phi_{s;\sI}^{\dagger *}(x)$ to better explore the nature of the error.
We first analyze the error characteristics of $\phi_{s;\sI}^{\dagger *}(x)$,
and finally demonstrate that these characteristics are fundamentally similar to those of the original scaled wavefront $\phi_{s;\sI}^{\dagger}(x)$.

By expanding $\phi(x+s)$ in the right-hand side of \EQ{eq:difference} using a Taylor series,
integrating both sides, and rearranging terms,
we obtain an expression for $\phi(x)$ in terms of the differential wavefront:
\begin{align}
  \label{eq:phi(x)}
  \phi(x) = \frac{1}{s}\int \fM(x,s)\,dx
  - \sum_{n=1}^\infty \frac{s^{n}}{(n+1)!}\phi^{(n)}(x).
\end{align}
When the shear amount $s$ is non-integer,
the FT-based reconstruction techniques,
such as the natural extension and spectral interpolation methods,
cannot directly handle the fractional shear.
In such cases, $s$ is typically replaced by a nearby integer shear $\sI$,
which inevitably introduces approximation errors in the reconstructed wavefront.

The difference between the fractional shear and the integer shear is denoted by $\ds$ as
\begin{align}
  \ds\equiv S-s.
\end{align}
When the wavefront is reconstructed using the integer shear
$\sI$, the observed quantity $\fM(x,s)$
still depends on the original fractional shear $s$.
However, in \EQ{eq:phi(x)}, both $s$ in the factor in
the first term and $s$ in the second term are replaced by
$\sI$. Thus, the wavefront is reconstructed as follows:
\begin{align}
\label{eq:phi'(x)}
\phi_{s;\sI}(x)=\frac{1}{\sI}\int \fM(x,s)\,dx
-\sum_{n=1}^\infty \frac{\sI^{n}}{(n+1)!}\phi_{s;\sI}^{(n)}(x).
\end{align}
For both \EQtwo{eq:phi(x)}{eq:phi'(x)},
if the second term on the right-hand sides are sufficiently
smaller than the first term, then dominant contribution comes from
the integral.
From the comparisons of the first terms, 
we obtain the relation between $\phi(x)$ and $\phi_{s;\sI}(x)$ given by 
\begin{align}
  \label{eq:def_phidag}
  \phi(x) \simeq \frac{\sI}{s}\phi_{s;\sI}(x) \equiv \phi_{s;\sI}^{\dagger}(x).
\end{align}
This expression indicates that the reconstructed wavefront $\phi_{s;\sI}(x)$
using the integer shear $S$
instead of the fractional shear $s$ that should originally be used for evaluation
is approximately equal to $\phi(x)$ scaled by a factor of $s/\sI$.
For instance, if $s=1.5$ and $\sI=1$,
the resulting error would be about 50\%,
which is generally unacceptable.
In other words, 
the scaled wavefront $\phi_{s;\sI}^\dagger(x)$
can reduce
the error in the first integral in \EQ{eq:phi'(x)}.
The residual error, $d_{s;\sI}^{\dagger}(x)$ in this approximation for $\phi(x)$
defined by
\begin{align}
   \label{eq:residual_err}
   d_{s;\sI}^{\dagger}(x)\equiv\phi_{s;\sI}^{\dagger}(x)-\phi(x)
\end{align}
will be evaluated later; however,
its exact expression is expected to be complicated,
involving nested summations.
To make this more practical, an approximated form
$\phi_{s;\sI}^{\dagger *}(x)$ instead of $\phi_{s;\sI}^{\dagger}(x)$
will be introduced, and its error will be analyzed.
The $n$-th derivatives of $\phi(x)$ are approximated as 
$\phi^{(n)}(x)\simeq\phi_{s;\sI}^{\dagger (n)}(x)$ from \EQ{eq:def_phidag}.
By using this approximation, 
the approximated phase $\phi_{s;\sI}^{\dagger *}(x)$ is defined
from $\phi_{s;\sI}^{\dagger}(x)$ as follows.
\begin{align}
  \label{eq:phidag(x)}
  \phi_{s;\sI}^{\dagger}(x)
   &=
   \frac{1}{s}\int \fM(x,s)\,dx
            -\sum_{n=1}^\infty \frac{\sI^{n}}{(n+1)!}\frac{\sI}{s}\phi_{s;\sI}^{(n)}(x)
   \nonumber\\
   &=
   \frac{1}{s}\int \fM(x,s)\,dx
            -\sum_{n=1}^\infty \frac{\sI^{n}}{(n+1)!}\phi_{s;\sI}^{\dagger\,(n)}(x)
   \nonumber\\
   &\simeq
   \frac{1}{s}\int \fM(x,s)\,dx
            -\sum_{n=1}^\infty \frac{\sI^{n}}{(n+1)!}\phi^{(n)}(x)
   \nonumber\\
   &\equiv
    \phi_{s;\sI}^{\dagger*}(x)
   .
\end{align}
Comparing $\phi_{s;\sI}^{\dagger*}(x)$
with $\phi(x)$
in \EQ{eq:phi(x)},
the only difference is that $s$
in the second term is replaced with $\sI$.
The difference is evaluated as
\begin{align}
  \label{eq:def-err-phidag}
  d_{s;\sI}^{\dagger*}(x)
    &
    \equiv\phi_{s;\sI}^{\dagger*}(x)-\phi(x)
   \nonumber\\&
    =
   \sum_{n=1}^{\infty}\frac{-1}{(n+1)!}\left(\sI^n-s^n\right)\phi^{(n)}(x)
   \nonumber\\&
    =
   \sum_{n=1}^{\infty}\frac{-1}{(n+1)!}\left((1+\hatds)^n-1\right)\hatphi^{(n)}(x),
  \\
  & \hatds\equiv\frac{\ds}{s}=\frac{\sI-s}{s},
   \nonumber\\
  &
   \label{eq:differential-hat}
   \hatphi^{(n)}(x)\equiv\frac{d^n\phi(x)}{d(x/s)^n}=s^n\phi^{(n)}(x),
\end{align}
where $\hatds$
and `$\hat{\phantom{\ds}}$'
in $\hatphi^{(n)}(x)$ represent the quantities
by which the $x$-axis is normalized by $s$; 
$\hatphi^{(n)}(x)$ has the same dimension as $\hatphi(x)$.
The factor 
$(1+\hatds)^n-1$ on the right-hand side of \EQ{eq:def-err-phidag}
can be expressed by a binomial theorem expansion as
\begin{align}
  \label{eq:binomial}
  (1+\hatds)^n-1
    =\sum_{m=1}^n \frac{n!}{m!(n-m)!}\hatds^{m}.
\end{align}
The summation of infinite terms of $n$ on the 
right-hand side in \EQ{eq:def-err-phidag} 
and the summation of the finite terms of $m$ from 
the binomial expansion is expressed as a double summation of $n$ and $m$.
If we exchange the order of the loop integers
$n$ and $m$ in the double summation and put $l=n-m+1$ $(n=l+m-1)$,
\begin{align}
   \sum_{n=1}^\infty &\sum_{m=1}^n a_n b_m
  =\sum_{m=1}^\infty \sum_{n=m}^\infty a_n b_m
   \nonumber\\&
  =\sum_{m=1}^\infty \sum_{l=1}^\infty a_{l-m+1} b_m
  =\smashoperator{\sum_{(m,l)\inN^2}}a_{l-m+1} b_m.
\end{align}
Thus, the domains of $m$ and $l$ are all natural numbers.
Therefore, $d_{s;\sI}^{\dagger*}(x)$ is
\begin{align}
  \label{eq:err-phidag}
  d_{s;\sI}^{\dagger*}(x)&=
   \sum_{(m,l)\inN^2}\alpha_{l,m}^* \hatphi^{(l+m-1)}(x)\,\hatds^m
   \nonumber\\&
     =
   \sum_{m\inN}\hatA_m^*(x)\,\hatds^m.
\end{align}
where
\begin{align}
   \label{eq:def-alpha}
   \alpha_{l,m}^*&\equiv \frac{-1}{(l+m)!\,m!\,({l-1})!}
   ,
   \\
  \label{eq:der_hatA}
   \hatA_m^*(x)&\equiv \sum_{l\inN}\alpha_{l,m}^* \hatphi^{(l+m-1)}(x)
   .
\end{align}
\EQ{eq:err-phidag} shows that  
the error of $\phi_{s;\sI}^{\dagger*}(x)$
is given by a power series of $\hatds$.
Except for special cases ($|s|<\dsamp/2$),
$|\hatds|\le 1/2$, and since $|\hatds|>|\hatds|^m$ ($m>1$),
this dominant term is the first-order term $\hatA_1^*(x)\hatds$,
which is proportional to $\hatds$.
Furthermore, if we assume that the term containing
the first derivative $\phi^{(1)}(x)$ is dominant among the terms of 
$\hatA_1^*(x)$ defined by \EQ{eq:der_hatA}, then
the magnitude of the error
is reduced as 
$|d_{s;\sI}^{\dagger*}(x)|\sim|\hatds\hatphi^{(1)}(x)/2|=|\ds\phi^{(1)}(x)/2|$,
where the dependence on $s$ does not appear explicitly.
In other words, under these conditions,
when choosing $\sI$ in the vicinity of $s$, for
any $s$, the error is proportional only to $\ds$, which is $s-\sI$.

The aforementioned $\phi_{s;\sI}^{\dagger*}(x)$ 
is the approximation of $\phi_{s;\sI}^{\dagger}(x)$ in order to roughly evaluate the nature of error.
In the remaining  part of this subsection,
we will evaluate the error of $\phi_{s;\sI}^{\dagger}(x)$ more precisely.
In general, the error increases as $\hatds$ increases; therefore, 
it is natural to assume that $d_{s;\sI}^{\dagger}(x)$ is 
also given by the power series expansion of $\hatds$, 
similar to $d_{s;\sI}^{\dagger*}(x)$ in \EQ{eq:err-phidag}.
However, since we cannot reject the possibility that uniform errors remain,
we include $\hatds^0$ in the expansion.
\begin{align}
  \label{eq:d-dag-0,1,2...}
  d_{s;\sI}^{\dagger}(x)=\smashoperator{\sum_{m\in\{0,1,2,\cdots\}}}\hatA_m(x) \hatds^m.
\end{align}
The difference between
\EQtwo{eq:phi'(x)}{eq:phi(x)}
is only the difference
between the second terms of each equation, which are
$\sI^n\phi_{s;\sI}^{\dagger\,(n)}(x)$ and 
$s^n\phi^{(n)}(x)$.
The difference of them is expressed as follows:
\begin{align}
  \label{eq:err-dagger2-term2}
  \sI^n&\phi_{s;\sI}^{\dagger\,(n)}
    -s^n\phi^{(n)}(x)
   \nonumber\\&
  =(1+\hatds)^n\left[\hatphi^{(n)}(x)+\hatd_{s;\sI}^{\dagger(n)}(x)\right]-\hatphi^{(n)}
  \nonumber\\&
  =\left((1+\hatds)^n-1\right)\left[\hatphi^{(n)}(x)+\hatd_{s;\sI}^{\dagger(n)
}(x)\right]+\hatd_{s;\sI}^{\dagger(n)}(x).
\end{align}
Then, the factor of the first term on the right-hand side
can be transformed in the same way as
the derivation of \EQ{eq:def-err-phidag} to \EQ{eq:err-phidag},
and the difference $\hatd_{s;\sI}^{\dagger}(x)$ becomes
\begin{align}
  \label{eq:err-dagger2}
  d_{s;\sI}^{\dagger}(x)
    &
    =
   \smashoperator{\sum_{(m,l)\inN^2}}\alpha_{l,m}^*\left[\hatphi^{(l+m-1)}(x)+\hatd_{s;\sI}^{\dagger(l+m-1)}(x)\right]\hatds^m
   \nonumber\\&
   +\sum_{n\inN}\beta_n \hatd_{s;\sI}^{\dagger(n)}(x),\\
   \beta_n&\equiv\frac{-1}{(n+1)!}.
\end{align}
The difference between $d_{s;\sI}^{\dagger}(x)$ in \EQ{eq:err-dagger2} and
$d_{s;\sI}^{\dagger*}(x)$ in \EQ{eq:err-phidag} is 
whether the derivatives of $\hatd_{s;\sI}^{\dagger}(x)$ are included .
To evaluate $\hatds$-dependency of $d_{s;\sI}^{\dagger}(x)$, 
we must evaluate the dominant term of the right-hand side.
The dominant term of the first term is the first order of $\hatds$ because $|\hatds|<1$.
However, the dominant term in the last term is unknown 
from the form of \EQ{eq:err-dagger2} because the nature 
of the derivatives, $d_{s;\sI}^{\dagger\,(n)}(x)$ is unknown.
The derivatives are expressed as a recursive form by taking the differentiation of \EQ{eq:err-dagger2}.
\begin{align}
  \label{eq:der_err-dagger2}
  \hatd_{s;\sI}^{\dagger\,(p)}(x)
    &=
   \nonumber\\&
   \smashoperator{\sum_{(m,l)\inN^2}}\alpha_{l,m}^*\left[\hatphi^{(p+l+m-1)}(x)+\hatd_{s;\sI}^{\dagger(p+l+m-1)}(x)\right]\hatds^m
   \nonumber\\&\hspace*{-1em}%
   +\sum_{n\inN}\beta_n \hatd_{s;\sI}^{\dagger(p+n)}(x),
\end{align}
where we applied $\hatd_{s;\sI}^{\dagger(0)}(x)=d_{s;\sI}^\dagger(x)$ 
from \EQ{eq:differential-hat} with $p=0$.
To examine the nature of $d_{s;\sI}^{\dagger}(x)$,
we first evaluate the zeroth-order term of $\hatds$, 
since it may become a larger term than the first-order term.
The minimum order of $\hatds$ in the first term on the 
right-hand side of \EQ{eq:err-dagger2} is the first order of
$\hatds$ as mentioned above; therefore,
the zeroth order is only included in higher-order derivatives of 
$d_{s;\sI}^{\dagger}(x)$ at the last term.
The derivative shown in \EQ{eq:der_err-dagger2} 
also includes the zeroth-order term only at the last term.
When we evaluate it sequentially, we obtain the zeroth-order term $\hatA_0(x)$ as
\begin{align}
  \hatA_0(x)
     &=\smashoperator{\sum_{(n,n',n'',\cdots)\inN^\infty}}
     \beta_{n}\beta_{n'}\beta_{n''}\cdots
     \hatd_{s;\sI}^{\dagger\,(n+n'+n''+\cdots)}(x)
    \nonumber\\&
    =\displaystyle\sum_{\Vn\inN^\infty}
     \left(\prod_{i\inN}\beta_{n_i}\right)
     \hatd_{s;\sI}^{\dagger\,(\sum_i n_i)}(x)
    .
\end{align}
\MEMO{Remove the additional numbering in Eq.(28).\\}%
If the derivative of $\hatd_{s;\sI}^{\dagger\,(\infty)}(x)$ is finite,
then $\hatA_0(x)=0$ because
$\beta_{n_i}<1$ and the factor $\prod_{i\inN}\beta_{n_i}=0$.
In other words, 
the minimum order of $d_{s;\sI}^{\dagger}(x)$ is the first order,
as well as $d_{s;\sI}^{\dagger*}(x)$.
Therefore, 
the power series expression of $d_{s;\sI}^{\dagger}(x)$ shown in \EQ{eq:d-dag-0,1,2...} is modified as
\begin{align}
  \label{eq:d_dagger_pow_series}
  d_{s;\sI}^{\dagger}(x)=\sum_{m\inN} \hatA_m(x) \hatds^m.
\end{align}
It should be noticed that $\hatA_m(x)$ is independent of $\hatds$ or $S-s$.
By evaluating the derivative of this equation with respect to $x$, we obtain
\begin{align}
  \label{eq:d_dagger_der}
  \hatd_{s;\sI}^{\dagger(p)}(x)=\sum_{m\inN} \hatA_m^{(p)}(x) \hatds^m.
\end{align}
The minimum order of $\hatds$ is also 1 for the derivatives.
Substituting this equation into \EQ{eq:err-dagger2} and expressing
the derivatives with summation by \EQ{eq:der_hatA}, 
we obtain the following form.
\begin{align}
  \hatd_{s;\sI}^{\dagger}(x)
   =&
   \smashoperator{\sum_{m\inN}}\hatA_m\hatds^m
       +\smashoperator{\sum_{(m,l,m')\inN^3}}\alpha_{l,m}^*\hatA_{m'}^{(l+m-1)}(x)\hatds^{m+m'}
   \nonumber\\&
   +\sum_{(n,m)\inN^3}\beta_n \hatA_m^{(n)}(x)\hatds^{m}.
\end{align}
Let us consider the $\hatds$-dependency from this form.
Regarding the first-order coefficient function, $\hatA_1(x)$, since $m+m'>1$, 
the second term on the right-hand side does not include
the first-order of $\hatds$; the terms including the first order are found in the first and the last terms. 
\begin{align}
  \hatA_1(x)=&\hatA_1^*(x)
    +\sum_{n\inN}
        \beta_n \hatA_1^{(n)}(x).
\end{align}
Taking a derivative of this equation to obtain a derivatives, 
and substituting the obtained derivative into the second term on the 
right-hand side of the above equation,
$\hatA_1^{(n)}(x)$ in the second term on the right-hand side is 
replaced by $\hatA_1^{*\,(n)}(x)$ and the additional term with a double summation as
\begin{align}
  \hatA_1(x)=&\hatA_1^*(x)
    +\sum_{n\inN}
        \beta_n \hatA_1^{*\,(n)}(x)
    \nonumber\\&
    +\sum_{(n,n')\inN}
        \beta_n\beta_{n'}\hatA_1^{(n+n')}(x).
\end{align}
Similarly,
by repeating this procedure, triple, quadruple, and more multiple summations appear in the last term.
The order of derivatives in the terms added to the last 
term is higher than that of the preceding terms.
Since the minimum differential order of
$\hatA_1^{*\,(n)}(x)$ is $n+1$ due to 
\EQ{eq:der_hatA},
the first term on the right-hand side is dominant,
if it can be assumed that
the first-order derivative is smaller than
the higher-order derivatives.
Therefore, $\hatA_1(x)\simeq\hatA_1^*(x)$, then
the properties of $d_{s;\sI}^{\dagger}(x)$
and $d_{s;\sI}^{\dagger*}(x)$ are the same.
The above analysis shows that, for fractional shear measurements,
the reconstruction error of the single-scaled wavefront is dominated
by the first-order term with respect to the shear deviation.
In practice, the fractional shear is rounded to the nearest integer value,
which implies that the normalized fractional shear deviation satisfies
$|\hat{\varepsilon}| < 0.5$.
Within this range, the reconstruction error increases monotonically
as $|\hat{\varepsilon}|$ increases,
indicating that the first-order error term constitutes the dominant
contribution to the overall reconstruction error,
while higher-order terms remain comparatively small.
This observation provides the primary motivation for the following subsection,
where a weighted integer shear averaging strategy is introduced
to suppress the dominant first-order error component.

\subsection{Reconstruction using two integer shears}
\label{sec:3-2-int-shear-dagger}

In subsection~\ref{sec:3-1-int-shear-dagger},
we calculated $\phi_{s;\sI}(x)$ by selecting a nearby integer shear $\sI$
from the measured difference wavefront $\fM(x,s)$ 
with a non-integer shear $s$.
We then evaluated the error of the scaled $\phi_{s;\sI}^\dagger(x)$.
Since the calculation is analytical, it is possible to change $\sI$,
and the errors are different.
By choosing two $\sIn1$ 
and $\sIn2$ as $\phi_{s;\sI}^\dagger(x)$, 
we obtain two equations whose respective errors are expressed 
as power series expansions shown in \EQ{eq:d_dagger_pow_series}.
By combining these two equations, 
we can eliminate any single term in the power series expansion. If 
the term to be eliminated is the most dominant 
first-order term, the order of the remaining error will be reduced to the second order.
To evaluate the residual error,
the error model is expressed as a second-order power series expansion.
\begin{align}
  \label{eq:dif_phidag_q-O2}
  \phi_{s;S_q}^{\dagger}(x) -\phi(x)&=\hatA_1(x)\hatdsn{q} +\hatA_2(x)\hatdsn{q}^2
   \nonumber\\ &
   +O(\hatdsn{q}^3),
   \qquad(q=\{1,2\}).
\end{align}
By multiplying the equations for
$q=1$ and $q=2$ by $\hatdsn2$ and $\hatdsn1$, respectively,
and subtracting them, we can eliminate 
the terms containing $\hatA_1(x)$.
Rearranging the result,
we obtain an interpolated function between
$\phi_{s;\sIn1}^{\dagger}(x)$ and $\phi_{s;\sIn2}^{\dagger}(x)$
as 

\begin{align}
   &
  \label{eq:d_sIv2}
  \phi_{s;\sIv2}^{\dagger}(x)-\phi(x)
    = -\hatA_2(x)\hatdsn1\hatdsn2+O(\hatds_{1,2}^3),
  \\ &
  \phi_{s;\sIv2}^{\dagger}(x)\equiv w_{\sIv2}^{(1)}\phi_{s;\sIn1}^{\dagger}(x)+w_{\sIv2}^{(2)}\phi_{s;\sIn2}^{\dagger}(x),
  \\ &
  \label{eq:wgt_sIv2}
  \left(\begin{array}{c} w_{\sIv2}^{(1)}\\w_{\sIv2}^{(2)}\end{array}\right)
   \equiv \frac{1}{\hatdsn2-\hatdsn1}
  \left(\begin{array}{c} +\hatdsn2\\ -\hatdsn1\end{array}\right).
\end{align}

Let $\sIn1$ and $\sIn2$ be integer shears close to the fractional shear $s$,
and let $\sIn2-\sIn1=\dsn2-\dsn1=\dsamp$, 
where $\dsamp$ represents the sampling interval of the measured phase difference.
The dependency of $\hatds_1$ and $\hatds_2$ in the first term 
on the right-hand side of \EQ{eq:d_sIv2} is
\begin{align}
  \label{eq:err_phi12dag}
  -\hatdsn1\hatdsn2=-\frac{\dsn1}{s}\frac{\dsn1+\dsamp}{s}=\frac{1}{s^2}\left[-\left(\dsn1+\frac{\dsamp}{2}\right)^2+\frac{\dsamp^2}{4}\right].
\end{align}
We can see that the error is a quadratic function
and becomes 0 at $\dsn1=0,\dsamp$, 
and has the maximum at $\dsn1=-\dsamp/2$\ ($\dsn2=\dsamp/2$).
As for $s$-dependency, since the derivative of $\phi(x)$ contained
in $\hatA_2(x)$ is second or higher, if we assume that the terms containing
second-order derivatives are more dominant than the terms containing third
or higher ones.
Furthermore, 
by using $\hatphi^{(2)}(x)=s^2\phi^{(2)}(x)$ which is derived from \EQ{eq:differential-hat},
the factor $1/s^2$ in the right-hand side of \EQ{eq:err_phi12dag} is cancelled, and
the $s$-dependency appears only in the term of $\dsn1^2$.

\subsection{Reconstruction using three integer shears}
\label{sec:3-3-int-shear-dagger}

In the case of three integer shears $S_q$ ($q=\{1,2,3\}$),
we can reduce two dominant error terms related to the first and the second order of $\hatds$.
Similar to the case of two shears, the model of the error 
is represented by the power series up to the third order to evaluate the remaining error.
\begin{align}
  \label{eq:eps3}
  \phi^{\dagger}_{s;\sIn{q}}(x)-\phi(x)&=\hatA_1(x)\hatdsn{q}+\hatA_2(x)\hatdsn{q}^2+\hatA_3(x)\hatdsn{q}^3
  \nonumber \\ &
  +O(\hatdsn{q}^4), \qquad (q=\{1,2,3\}).
\end{align}
From these, eliminating terms including
$\hatA_1(x)$ and $\hatA_2(x)$, we obtain
\begin{align}
   &
  \label{eq:d_sIv3}
  \phi_{s;\sIv3}^{\dagger}(x)-\phi(x)
    =-\hatA_3(x)\hatdsn1\hatdsn2\hatdsn3+O(\hatdsn{1,2,3}^4),
    \\&
  \phi_{s;\sIv3}^{\dagger}(x)\equiv\sum_{q=1}^3 w_{\sIv3}^{(q)}\phi^{\dagger}_q,
    \\&
  \left(\begin{array}{c}
    w_{\sIv3}^{(1)}\\w_{\sIv3}^{(2)}\\w_{\sIv3}^{(3)}
  \end{array}\right)
    =
     \displaystyle\frac{-1}{\Delta^3_{\hatds}}
        \left(\begin{array}{c}
      \hatdsn2\hatdsn3(\hatdsn3-\hatdsn2)\\
      \hatdsn3\hatdsn1(\hatdsn1-\hatdsn3)\\
      \hatdsn1\hatdsn2(\hatdsn2-\hatdsn1)
    \end{array}\right),
    \\
    &
    \Delta^3_{\hatds}\equiv(\hatdsn3-\hatdsn2)(\hatdsn2-\hatdsn1)(\hatdsn1-\hatdsn3).
\end{align}

Let the integer shears $\sIn{q}$\ $(q=\left\{1,2,3\right\})$
be arranged in order at equal intervals of $\dsamp$.
\begin{align}
&
\sIn1+\dsamp=\sIn2=\sIn3-\dsamp,
\\
\label{eq:ds-relation}
&
\dss1+\dsamp=\dss2=\dss3-\dsamp.
\end{align}
In this case, the error factor $\hatdsn1\hatdsn2\hatdsn3$ is
\begin{align}
  \label{eq:err_phi123dag}
  -\hatdsn1\hatdsn2\hatdsn3=\frac{1}{s^3}\dss2(\dsamp^2-\dss2^2),
\end{align}
and is 0 at $\dss2=0,\pm\dsamp$,
and has the extreme values at $\dss{2}=\pm\dsamp/\sqrt{3}$.
As for the dependence on $s$, as in subsections~\ref{sec:3-1-int-shear-dagger} 
and~\ref{sec:3-2-int-shear-dagger}, if 
the term containing the third derivative is larger than
the term containing the fourth derivative, then
the factor $1/s^3$ is canceled, and the $s$-dependency appears only as $\ds_2^3$. 

After suppressing the dominant first- and second-order errors
using the proposed weighted integer shear averaging method,
the remaining reconstruction error is mainly governed by higher-order terms
with respect to the shear deviation.
In principle,
these higher-order error contributions can be further reduced
by increasing the number of integer shear reconstructions included in the averaging process.
The corresponding weighting coefficients can be obtained
by solving a system of simultaneous equations numerically.
In the present study,
three integer shear reconstructions were found to provide sufficient attenuation of the fractional-shear-induced error
while maintaining computational efficiency.
Therefore, further extension to higher-order formulations is not pursued in this work.
%

\section{Numerical Simulations}
\label{sec:results}

\begin{figure}[b] 
 \centering {
  \parbox{0.8\hsize}{%
    \includegraphics[width=\hsize]{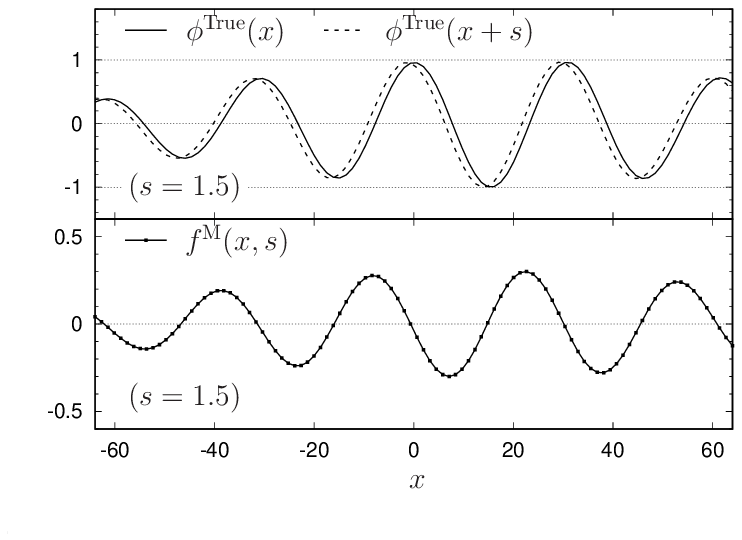}
  }%
 }
  \caption{The simulation data as
  true wavefront $\phi^{\rm True}(x)$ and differential phase $\fM(x,s)$ for $s = 1.5$.
  }
  \label{fig:input_simulation}
\end{figure}

\begin{figure} 
  \centering {
  \parbox{0.75\hsize}
  {%
    (a) $\phi_{s;\sI}(x)$ and $\Delta\phi_{s;\sI}(x)$\\
    \includegraphics[width=\hsize]{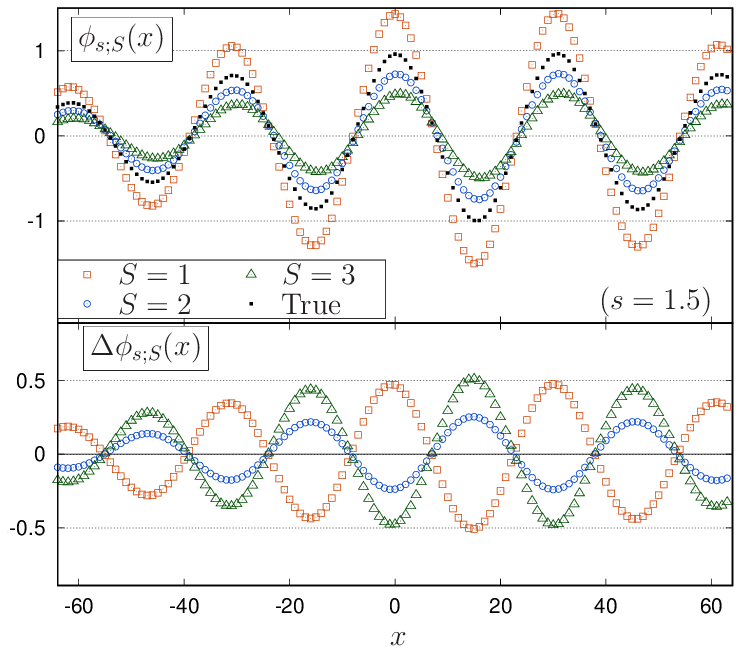}
  } \\%
  \parbox{0.75\hsize}
  {%
    (b) $\phi_{s;\sI}^\dagger(x)$ and $\Delta\phi_{s;\sI}^\dagger(x)$\\
     \includegraphics[width=\hsize]{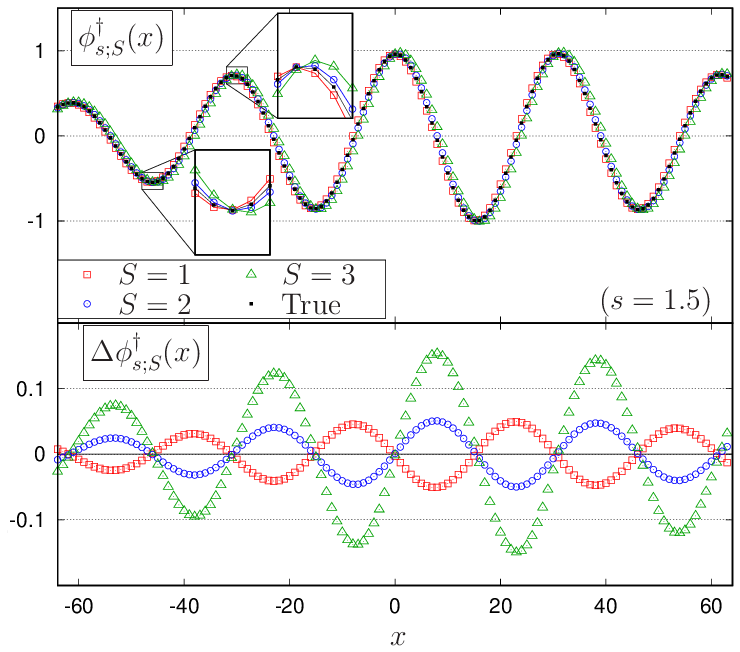}
  }%
  \caption{
     Comparisons of errors between the wavefront without and with scaling:
    (a) Reconstructed wavefronts by the basic method without scaling, $\phi_{s,\sI}(x)$
    and their errors, $\Delta\phi_{s;\sI}(x)\equiv\phi_{s,\sI}-\phi^{\rm True}(x)$.
    (b) Scaled wavefronts, $\phi_{s;\sI}^\dagger(x)=\frac{\sI}{s}\phi_{s;\sI}(x)$
    and their errors, $\Delta\phi_{s;\sI}^\dagger(x)$.
    Note: The errors range in the lower panels differ between subfigures (a) and (b).
  }
  \label{fig:err-single}
  }
\end{figure}

\begin{figure} 
\centering {
  \parbox{0.8\hsize}{%
    \includegraphics[width=\hsize]{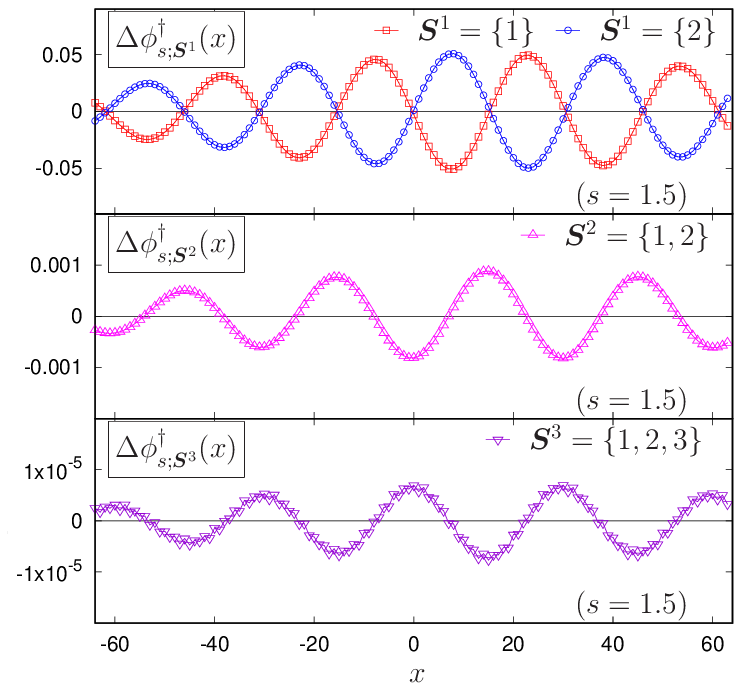}
  }%
  \caption{
    Errors by the weighted averaging method.
     The superscripts 1, 2, and 3 of $\sIv{}$ express the number of the scaled wavefronts
     employed in the weighted averaging method.
     Note: The vertical scales differ between the panels.
  }
  \label{fig:err-average}
  }
\end{figure}

To assess the applicability of the proposed method, 
we consider a representative case in which the true wavefront is known. 
This enables a quantitative evaluation of the reconstruction accuracy. 
In the numerical simulations,
a smooth  test wavefront is employed
in order to satisfy the applicability condition of the reconstruction method for integer shear \cite{Tomioka:23}
and to isolate the effects induced purely by fractional shear.

As a representative example, we selected 
the following phase modulated sinusoidally, 
which is considered to be a little more complicated than the actual object.
\begin{align}
  \label{eq:phi-example}
  \phi^{\rm True}(x)&=\exp{\left(-\frac{(x-x_c)^2}{w^2}\right)}\cos(k_0 x)\MODIFIED{,}{}
   \nonumber\\&\qquad
   (k_0=2\pi/L).
\end{align}
The true wavefront is specified by setting the parameters 
to $(x_c,w,L)=(16,50,30)$. 
Figure~\ref{fig:input_simulation} shows the true wavefront 
together with the simulated differential phase 
$\fM(x, s)$ for $s = 1.5$, where $\fM(x, s)$ is defined in \EQ{eq:difference}.
The dots on $\fM(x,s)$ are the discrete input data used in 
the reconstruction simulations, which serve as an alternative to 
the measured wavefront difference,
where the sampling intervals of them are $\dsamp=1$.
Below, we show the errors of the reconstructed 
wavefronts calculated by some methods
from $\fM(x;s)$
by a comparison with $\phi^{\rm True}(x)$.
The reconstructed wavefront contains an ambiguous uniform bias, called 
the piston term, that cannot be resolved from $\fM(x;s)$ alone.
Therefore, the difference is evaluated after subtracting the average of the wavefronts as
\begin{align}
  \Delta{\phi^{\rm cal}}(x)&\equiv\widetilde{\phi}^{\rm cal}(x)-\widetilde{\phi}^{\rm True}(x),
   \\
  \widetilde{\phi}(x)&=\phi(x)-\bar{\phi},\quad\bar{\phi}\equiv\Ave{\phi(x)},
\end{align}
where $\Ave{\ }$ and $\phi^{\rm cal}$ represent an average over $x$ and
a calculated wavefront by a certain method, respectively.

Figure~\ref{fig:err-single} presents the reconstructed wavefronts 
obtained from the simulated data using two approaches:  
(a) the basic method denoted as $\phi_{s;S}(x)$,
where the spectral interpolation method coupling with the natural extension method
\cite{Tomioka:23} was used as the basic method;
and (b) the scaled method with the scale factor $S/s$ shown in \EQ{eq:def_phidag}
employing a single integer shear, denoted as $\phi_{s;S}^\dagger(x)$.  
Reconstructions are shown for different integer shear values.  
The reconstructed wavefronts obtained 
by the basic method
show that the reconstruction for $S=1$ which is the case of $S<s$ exceeds 
the true wavefront $\phi^{\rm True}(x)$, 
shown as black dots in Fig.~\ref{fig:err-single}.
Whereas the reconstructions for the cases where 
$S>s$ ($S=2,3$) underestimate $\phi^{\rm True}(x)$.  
Comparing \EQtwo{eq:phi(x)}{eq:phi'(x)}, 
the main difference arises from the scaling factors of the first integral terms,
which are $1/s$ and $1/\sI$, respectively.
The higher-order correction terms on the right-hand sides are also scaled 
by powers of $s$ and $\sI$, leading to corresponding overestimations and 
underestimations in the reconstructed wavefront.
Therefore, these deviations are mutually consistent and primarily governed 
by the difference between $s$ and $\sI$.
In contrast, the reconstructions obtained by the scaled method based on 
\EQ{eq:def_phidag} exhibit close agreement with 
the true wavefront across all tested integer shear values.  
From the error perspective of $\Delta\phi_{s;S}^\dagger(x)$, the reconstructions for $S=1$ and $S=2$
yield nearly identical error magnitudes, whereas the reconstruction
for $S=3$ results in a visibly larger error.
This behaviour is consistent with the theoretical dependence on
the distance $\epsilon = |s - S|$ between the fractional shear $s = 1.5$
and the substituted integer shears: $\epsilon = 0.5$ for both $S=1$
and $S=2$, and $\epsilon = 1.5$ for $S=3$.
In addition, it exhibits
$\Delta\phi_{s;2}^\dagger(x)\simeq-\Delta\phi_{s;1}^\dagger(x)$.

Figure~\ref{fig:err-average} shows the error reduction by
using the proposed weighted averaging method, 
which employs multiple scaled wavefronts.
The number of employed wavefronts is represented as the superscript, $m$, of $\sIv{m}$.
Hereafter, we denote $\phi_{s;\sI}^\dagger(x)$ as 
$\phi_{s;\sIv1}^\dagger(x)$ because they are the same.
As mentioned above, the errors $\Delta\phi_{s;\sIv1}^{\dagger}(x)$ for
$S = 1$ and $S = 2$, exhibit opposite phases with almost the same amplitudes.
Since the weights in \EQ{eq:wgt_sIv2} are 
$w_{\sIv2}^{(1)}=w_{\sIv2}^{(2)}=1/2$ when $s=1.5$,
averaging these reconstructions effectively 
reduces the residual error
as demonstrated in $\Delta\phi_{s;\sIv2}^\dagger(x)$.
Furthermore, the use of three integer shear values provides additional 
suppression of higher-order components, 
resulting in lower overall error magnitudes compared to the single-shear case.

\begin{figure}[t]
 \centering {
  \parbox{0.45\hsize}{%
    (a)\\
    \includegraphics[width=\hsize]{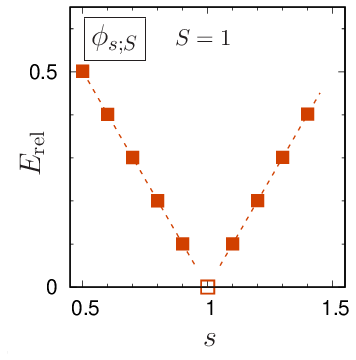}
  }%
  \parbox{0.45\hsize}{%
    (b)\\
    \includegraphics[width=\hsize]{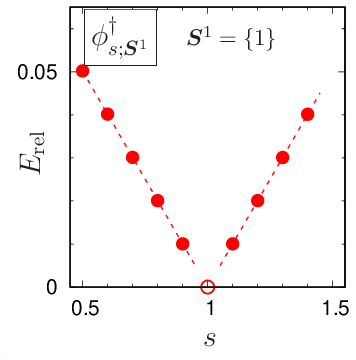}
  } \\%
  \parbox{0.45\hsize}{%
    (c)\\
    \includegraphics[width=\hsize]{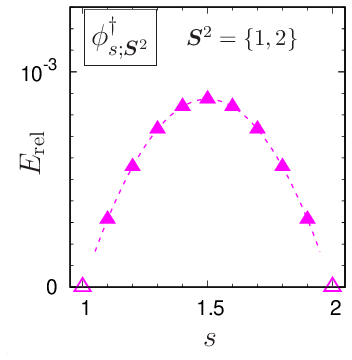}
  }%
  \parbox{0.45\hsize}{%
    (d)\\
    \includegraphics[width=\hsize]{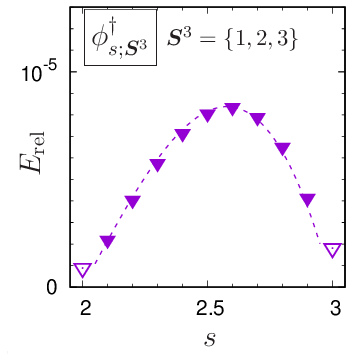}
  }%
  }
  \caption{Shear dependence of relative reconstruction errors.  
          (a) Reconstructed wavefront using the basic method.  
          (b)-(d) Reconstructed wavefronts evaluated by the proposed method 
	  using single, two, and three integer shear values, respectively.
	  Open and solid symbols denote integer and fractional shears, respectively.  
          The dashed lines in subfigures represent 
	  the theoretical error obtained by least-squares fitting 
	  for solid points with linear, 
	  quadratic, or cubic functions.  
          The scales of the vertical axes differ between subfigures.
          The interval between evaluations for $s$ is $\dsamp/10$.
  }
  \label{fig:shear_dependence_local}
\end{figure}

To further investigate the error nature of the proposed algorithm, 
we examine the dependence of the reconstruction error on the shear value $s$. 
The reconstructed wavefronts are compared for different shear values obtained using  
the basic method \cite{Tomioka:23} for a single shear and  
the proposed weighted average method, which employs single or multiple integer shears,
as shown in Fig.~\ref{fig:shear_dependence_local}.  
The reconstruction error, $E_{\rm rel}$, is evaluated using a relative error metric, 
defined as
\begin{align}
  &
  \Erel{\phi^{\rm cal}}\equiv\frac{\RMS{\Delta\phi^{\rm cal}(x)}}{\RMS{\widetilde{\phi}^{\rm True}(x)}},
  \\
  &
  \RMS{a(x)}\equiv\sqrt{\Ave{(a(x))^2}},
\end{align}
where  $a(x)$ represents an arbitrary function.
From Fig.~\ref{fig:shear_dependence_local},
we can see that the error decreases when the fractional shear
$s$ approaches  the integer shear $S$ in all cases.
The error dependencies of $s$ of $\phi_{s;\sIv{m}}^\dagger(x)$ for $m=\{1,2,3\}$
are well fitted to linear, quadratic, and cubic functions, respectively.
And the maximum errors are found at $s=1.5$ 
for $\phi_{s;\sIv2}(x)$ and $s=2.6$ 
for $\phi_{s;\sIv3}(x)$, which are corresponding 
to $\dss{1}=0.5$ and $\dss{2}=0.6\simeq 1/\sqrt{3}$, respectively.
These properties are consistent with the error estimation shown 
in subsections \ref{sec:3-1-int-shear-dagger}, 
\ref{sec:3-2-int-shear-dagger}, and \ref{sec:3-3-int-shear-dagger}.


\begin{figure}[t]
\centering {
  \parbox{0.95\hsize}{%
    \includegraphics[width=\hsize]{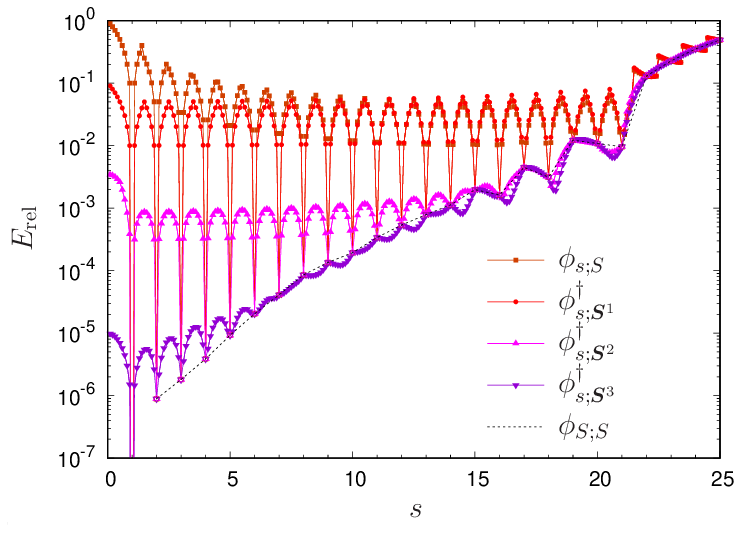}
  }%
  \caption{Shear dependence of relative errors.
   The open symbols at the local minima and solid symbols represent the case
   where $s$ are integer shears and fractional shears, respectively.
  }
\label{fig:shear}
  }
\end{figure}

Figure \ref{fig:shear} summarizes the shear dependence of the reconstruction errors 
$\Delta\phi_{s;\sI}(x)$, 
and $\Delta\phi_{s;\sIv{m}}^\dagger(x)$ for $m=\{1,2,3\}$.
The integer shears, $S$ and $\sIv{m}$, are selected as the nearest integer shear 
for each $s$ except $S=0$ because $\phi_{s;0}^\dagger(x)$ 
is identical to zero, having no additional information.
If $S=0$ is  one of the nearest integer shears, 
a negative shear is selected instead of $S=0$; e.g.,
$\sIv2=\{-1,1\}$ for $s\in(0,0.5)$, $\sIv3=\{-2,-1,1\}$ 
for $s\in(0,0.5)$, $\sIv3=\{-1,1,2\}$ for $s\in[0.5,2)$.
In Fig.\ref{fig:shear}, the open symbols connected with a dashed line do not include 
the contribution of the fractional shear; the error which 
increases with increasing $s$ is caused by the 
spectral interpolation method \cite{Tomioka:23}.
The contributions to the error of the fractional shear represent 
the difference between the dashed line
and solid symbols for each $s$.
We can find that the local maxima of errors due to the fractional shear 
are almost constant for each $\phi_{s;\sIv{m}}(x)$.
This nature is consistent with the nature shown in section \ref{sec:algorithm}.
The exceptions are found at $s<1$.
This reason is that the selected integer shears $S_q$ are not equally spaced 
at $s<1$ where the negative integer shear is selected, as mentioned above. %

In addition, if the norm of the normalized derivatives of $\hat{\phi}(x)$ satisfies,
\begin{align}
  \label{eq:der_cond}
  &
    \Norm{\hat{\phi}^{(m-1)}(x)}\gg\Norm{\hat{\phi}^{(m)}(x)} \qquad (p\inN),  \\
  &
  \Norm{\hatphi^{(m)}(x)}\equiv\left(\displaystyle\int |\hatphi^{(m)}(x)|^2\,dx\right)^{1/2},
\end{align}
we can estimate the amount of error for each method from
\EQthree{eq:d_dagger_pow_series}{eq:d_sIv2}{eq:d_sIv3}.
The RMSs, which are proportional to the norms,
of derivatives $\phi^{(m)}(x)=s^m\hatphi^{(m)}(x)$ evaluated numerically, 
were 0.58, 0.12, 0.024, and 0.0049 for $m=0,1,2,3$, respectively,
where the ratio between the norms of the adjacent order of 
differentials is 0.21, which almost equals $k_0$.
Thus, the model shown in \EQ{eq:phi-example}
satisfies \EQ{eq:der_cond} where $s\ll L$.
Under this condition,
the dominant error of
$\phi_{s;\sIv{m}}^{\dagger}(x)$ is induced by $\hatA_m(x)$.
Moreover, the approximation $\hatA_m^{*}(x)\approx\hatA_m(x)$ 
holds, and the dominant term of $\hatA_m(x)$ is determined 
by the minimum derivative of $\hatphi(x)$ as
\begin{align}
  \hatA_m(x)\sim \alpha_{1,m}^*\hatphi^{(m)}(x).
\end{align}
Therefore, the theoretical relative error in the weighted averaging methods is
\def\Epsilon{{\cal E}}
\begin{align}
  \Ereltheo{\phi_{s;\sIv{m}}^\dagger}&\sim
    |\alpha_{1,m}^*|\Epsilon_{\max}^m\frac{\RMS{\phi^{(m)}(x)}}{\RMS{\phi^{(0)}(x)}},
   \\
   \Epsilon_{\max}^m&=\max_s\left\{\left|\prod_{i=1}^m\epsilon_i\right|\right\}.
\end{align}
From the definition of $\alpha_{l,m}^*$ shown in \EQ{eq:def-alpha} 
$\alpha_{1,m}^*=\{1/4,1/12,1/144\}$ is obtained.
The factor of $\Epsilon_{\max}^m$ can be estimated
from the local maxima of the products in
\EQtwo{eq:err_phi12dag}{eq:err_phi123dag};
i.e., $|\ds|=\dsamp/2$ for $\phi_{s;\sIv1}^{\dagger}(x)$, 
$\phi_{s;\sIv2}^{\dagger}(x)$ and $\ds=\dsamp/\sqrt{3}$ for $\phi_{s;\sIv3}^{\dagger}(x)$;
and the results are $\Epsilon_{\max}^m=\{\dsamp/2,\dsamp^2/4,2\dsamp^3/(3\sqrt{3})\}$.
The theoretical errors are estimated from these quantities as
$\Ereltheo{\phi_{s;\sIv{m}}}=\{0.050,0.00086,2.3\times10^{-5}\}$ for $m=\{1,2,3\}$, respectively.
For comparisons,
the actual relative errors for the result shown in Fig.\ref{fig:shear} were
$\Erel{\phi_{s;\sIv{m}}^{\dagger}(x)}\simeq\{0.050,0.00088, 0.81\times10^{-5}\}$ around $s\simeq 2.5$,
which agree with the theoretical ones for $\sIv1$ and $\sIv2$, 
and the order is the same for $\sIv3$ in which the error for the integer shear is comparative.

In this section, we demonstrated that the errors of
$\phi_{s;\sIv{m}}^\dagger(x)$ are, 
qualitatively and quantitatively, 
consistent with the theory shown in section~\ref{sec:algorithm}.
However, the readers may wonder whether this result
was because
the phase model shown in \EQ{eq:phi-example} just happens to satisfy \EQ{eq:der_cond}.
The validity of \EQ{eq:der_cond} follows from the spectral characteristics of
$\hat{\phi}(x)$.
Since the derivative in the Fourier domain is weighted by $|k|$,
if the reconstructed wavefront is sufficiently smooth, i.e.,
the Fourier spectrum of $\hatphi(x)$ is effectively limited as $|k|\lesssim k_{\max}$,
the norm of the $m$-th derivative of $\hatphi(x)$ satisfies
\begin{align}
  \Norm{\hatphi^{(m+1)}(x)}
  &\lesssim s k_{\max}\Norm{\hatphi^{(m)}(x)}.
\end{align}
The effective limit of the model shown in \EQ{eq:phi-example} was $k_{\max}\simeq k_0$,
which is a result from the convolution theorem
in which the peaks represented by a Dirac's delta function, $\delta(k\pm k_0)$, in the spectral 
domain are blurred by the narrow Gaussian function with the spectral width $1/w$.
And this model satisfies, \EQ{eq:der_cond} as mentioned above.
In the case of a more practical model
such as a superposition of several Gaussian functions with the minimum width $w_{\min}$
without cosine modulation ($k_0=0$),
the effective limit $k_{\max}$ is determined only from 
the spectral width of the Gaussian functions as $k_{\max}\approx 1/w_{\min}$.
This suggests that the condition of \EQ{eq:der_cond} 
is met even when $s$ is larger, since $k_{\max}$ is smaller than the cosine-modulated model.
Consequently, it can be concluded that the condition to reduce the error 
due to the fractional shear is
\begin{align}
  s k_{\max}\ll1,
\end{align}
and the weighted averaging method is applicable for models without steeply
changing distributions, such as step-shaped phase distributions.
Furthermore, the proposed weighting strategy relies on accurate knowledge
of the normalized shear deviation.
Small calibration errors in the shear amount affect
the weighting coefficients only to the first order
and therefore do not significantly degrade the reconstruction accuracy
within the applicable range. 
In the numerical simulations,
a smooth  test wavefront is employed
in order to satisfy the applicability condition of the reconstruction framework
and to isolate the effects induced purely by fractional shear.
This choice is not incidental but is justified
by the theoretical discussions presented in the latter part of this section,
where the validity of \EQ{eq:der_cond} is shown to follow
from the spectral characteristics of the reconstructed wavefront.
However, regarding noise, it should be emphasized that
the proposed method is not intended to suppress random noise with a white spectrum.
Such noise contains high spatial-frequency components
that lie outside the applicability condition of the method.
Consequently, direct noise reduction cannot be expected for white-spectrum noise
using the proposed reconstruction framework.
To apply the proposed method for the measurement with thermal noise,
the effective noise level should be reduced through repeated measurements and averaging.

\section{Conclusion}
\label{sec:conclusion}

Wavefront reconstruction in lateral shearing interferometry is 
often hindered by the practical difficulty of realizing shears that 
are integer multiples of the detector sampling interval. 
Conventional methods that approximate fractional shears by 
the nearest integer introduce systematic reconstruction errors, 
which can become significant in precision applications. 
In this work, we proposed a weighted integer shear averaging 
method that systematically reduces these errors 
by combining reconstructions from multiple nearby integer shears 
with analytically chosen weights.
Theoretical error analysis demonstrated that 
two-shear averaging cancels first-order error terms, while 
three-shear averaging further suppresses second-order contributions. 
Numerical simulations confirmed these predictions, showing that 
the proposed method achieves substantially 
lower RMS error than conventional single-shear reconstruction. 
The approach is computationally efficient, 
requires no changes to existing shear interferometer hardware, 
and can be naturally extended to two-dimensional cases 
by cooperating with the other method to solve the piston term problem.
Overall, the results indicate that weighted integer shear averaging 
provides a practical and accurate strategy for wavefront reconstruction under 
fractional shear conditions.
It should be emphasized that the proposed method operates purely at the reconstruction level and is independent of specific optical configurations. The dominant contribution of this study lies in the closed-form theoretical error analysis and analytical derivation of optimal weights.
Including experimental validation would require system-specific calibration, fringe processing, and noise modeling, which would broaden the scope of the present paper and obscure the central theoretical contribution. Experimental verification is, therefore, left as a topic for future work.
%


\section*{Acknowledgements}

This research was supported by Japan Society for the Promotion of Science (JSPS), 22K04117.



\begin{thebibliography}{10}
\expandafter\ifx\csname url\endcsname\relax
  \def\url#1{\texttt{#1}}\fi
\expandafter\ifx\csname urlprefix\endcsname\relax\def\urlprefix{URL }\fi
\expandafter\ifx\csname href\endcsname\relax
  \def\href#1#2{#2} \def\path#1{#1}\fi

\bibitem{Yoshizawa2015}
T.~Yoshizawa (Ed.), Handbook of Optical Metrology: Principles and Applications,
  2nd Edition, CRC Press, 2015.

\bibitem{Sirohi2009}
R.~S. Sirohi, Introduction to Optical Metrology, 2nd Edition, Taylor \&
  Francis, 2009.

\bibitem{Schnars2015}
U.~Schnars, C.~Falldorf, J.~Watson, W.~J\"{u}ptner, Digital Holography and
  Wavefront Sensing: Principles, Techniques and Applications, Springer, 2015.

\bibitem{Mana:18}
G.~Mana, E.~Massa, C.~P.~S. Sasso, Wavefront errors in a two-beam
  interferometer, Metrologia 55~(4) (2018) 535--540.
\newblock \href {https://doi.org/10.1088/1681-7575/aacae6}
  {\path{doi:10.1088/1681-7575/aacae6}}.

\bibitem{Zhang:11}
S.~Zhang, Z.~Xu, B.~Chen, L.~Yan, J.~Xie, The comparison of environmental
  effects on michelson and fabry--perot interferometers utilized for the
  displacement measurement, Sensors 10~(4) (2011) 2577--2593.
\newblock \href {https://doi.org/10.3390/s100402577}
  {\path{doi:10.3390/s100402577}}.

\bibitem{Bon:09}
P.~Bon, G.~Maucort, B.~Wattellier, S.~Monneret, Quadriwave lateral shearing
  interferometry for quantitative phase microscopy of living cells, Opt.
  Express 17~(15) (2009) 13080--13094.
\newblock \href {https://doi.org/10.1364/OE.17.013080}
  {\path{doi:10.1364/OE.17.013080}}.

\bibitem{Tomioka:15}
S.~Tomioka, S.~Nishiyama, S.~Heshmat, Y.~Hashimoto, K.~Kurita,
  {Three-dimensional gas temperature measurements by computed tomography with
  incident angle variable interferometer}, in: C.~A. Bouman, K.~D. Sauer
  (Eds.), Computational Imaging XIII, Vol. 9401, International Society for
  Optics and Photonics, SPIE, 2015, p. 94010J.
\newblock \href {https://doi.org/10.1117/12.2082499}
  {\path{doi:10.1117/12.2082499}}.

\bibitem{Tomioka:17}
S.~Tomioka, S.~Nishiyama, N.~Miyamoto, D.~Kando, S.~Heshmat, Weighted
  reconstruction of three-dimensional refractive index in interferometric
  tomography, Appl. Opt. 56~(24) (2017) 6755--6764.
\newblock \href {https://doi.org/10.1364/AO.56.006755}
  {\path{doi:10.1364/AO.56.006755}}.

\bibitem{Ling:17}
T.~Ling, J.~Jiang, R.~Zhang, Y.~Yang, Quadriwave lateral shearing
  interferometric microscopy with wideband sensitivity enhancement for
  quantitative phase imaging in real time, Scientific reports 7~(1) (2017)
  1--14.
\newblock \href {https://doi.org/10.1038/s41598-017-00053-7}
  {\path{doi:10.1038/s41598-017-00053-7}}.

\bibitem{Murty:64}
M.~V. R.~K. Murty, The use of a single plane parallel plate as a lateral
  shearing interferometer with a visible gas laser source, Appl. Opt. 3~(4)
  (1964) 531--534.
\newblock \href {https://doi.org/10.1364/AO.3.000531}
  {\path{doi:10.1364/AO.3.000531}}.

\bibitem{Riley:77}
M.~E. Riley, M.~A. Gusinow, Laser beam divergence utilizing a lateral shearing
  interferometer, Appl. Opt. 16~(10) (1977) 2753--2756.
\newblock \href {https://doi.org/10.1364/AO.16.002753}
  {\path{doi:10.1364/AO.16.002753}}.

\bibitem{Rimmer:75}
M.~P. Rimmer, J.~C. Wyant, Evaluation of large aberrations using a
  lateral-shear interferometer having variable shear, Appl. Opt. 14~(1) (1975)
  142--150.
\newblock \href {https://doi.org/10.1364/AO.14.000142}
  {\path{doi:10.1364/AO.14.000142}}.

\bibitem{Harbers:96}
G.~Harbers, P.~J. Kunst, G.~W.~R. Leibbrandt, Analysis of lateral shearing
  interferograms by use of zernike polynomials, Appl. Opt. 35~(31) (1996)
  6162--6172.
\newblock \href {https://doi.org/10.1364/AO.35.006162}
  {\path{doi:10.1364/AO.35.006162}}.

\bibitem{Okuda:00}
S.~Okuda, T.~Nomura, K.~Kamiya, H.~Miyashiro, K.~Yoshikawa, H.~Tashiro,
  High-precision analysis of a lateral shearing interferogram by use of the
  integration method and polynomials, Appl. Opt. 39~(28) (2000) 5179--5186.
\newblock \href {https://doi.org/10.1364/AO.39.005179}
  {\path{doi:10.1364/AO.39.005179}}.

\bibitem{Dai:13}
F.~Dai, F.~Tang, X.~Wang, O.~Sasaki, M.~Zhang, High spatial resolution zonal
  wavefront reconstruction with improved initial value determination scheme for
  lateral shearing interferometry, Appl. Opt. 52~(17) (2013) 3946--3956.
\newblock \href {https://doi.org/10.1364/AO.52.003946}
  {\path{doi:10.1364/AO.52.003946}}.

\bibitem{Mochi:15}
I.~Mochi, K.~A. Goldberg, Modal wavefront reconstruction from its gradient,
  Appl. Opt. 54~(12) (2015) 3780--3785.
\newblock \href {https://doi.org/10.1364/AO.54.003780}
  {\path{doi:10.1364/AO.54.003780}}.

\bibitem{Ling:15}
T.~Ling, Y.~Yang, D.~Liu, X.~Yue, J.~Jiang, J.~Bai, Y.~Shen, General
  measurement of optical system aberrations with a continuously variable
  lateral shear ratio by a randomly encoded hybrid grating, Appl. Opt. 54~(30)
  (2015) 8913--8920.
\newblock \href {https://doi.org/10.1364/AO.54.008913}
  {\path{doi:10.1364/AO.54.008913}}.

\bibitem{Shi:23}
R.~Shi, H.~Liu, Y.~Shao, J.~Bai, Wavefront reconstruction for double-grating
  ronchi lateral shearing interferometry with nonlinear optimization, in: SPIE
  Optical Design and Testing XIII (Proceedings of SPIE Vol. 12765), 2023, p.
  127650S.
\newblock \href {https://doi.org/10.1117/12.2688489}
  {\path{doi:10.1117/12.2688489}}.

\bibitem{Luo:25}
Y.~Luo, Z.~Luo, D.~Hou, D.~Bian, Y.~He, W.~Jiang, Y.~Li, High-flexibility
  single-shot wavefront measurement with dual-lateral shearing interferometry,
  Optics and Lasers in Engineering 186 (2025) 108792.
\newblock \href {https://doi.org/10.1016/j.optlaseng.2024.108792}
  {\path{doi:10.1016/j.optlaseng.2024.108792}}.

\bibitem{Southwell:80}
W.~Southwell, Wave-front estimation from wave-front slope measurements, J. Opt.
  Soc. Am. 70~(8) (1980) 998--1006.
\newblock \href {https://doi.org/10.1364/JOSA.70.000998}
  {\path{doi:10.1364/JOSA.70.000998}}.

\bibitem{Zou:00}
W.~Zou, Z.~Zhang, Generalized wave-front reconstruction algorithm applied in a
  shack--hartmann test, Appl. Opt. 39~(2) (2000) 250--268.
\newblock \href {https://doi.org/10.1364/AO.39.000250}
  {\path{doi:10.1364/AO.39.000250}}.

\bibitem{Dai:12}
F.~Dai, F.~Tang, X.~Wang, O.~Sasaki, Generalized zonal wavefront reconstruction
  for high spatial resolution in lateral shearing interferometry, J. Opt. Soc.
  Am. A 29~(9) (2012) 2038--2047.
\newblock \href {https://doi.org/10.1364/JOSAA.29.002038}
  {\path{doi:10.1364/JOSAA.29.002038}}.

\bibitem{Dai:16}
F.~Dai, J.~Li, X.~Wang, Y.~Bu, Exact two-dimensional zonal wavefront
  reconstruction with high spatial resolution in lateral shearing
  interferometry, Optics Communications 367 (2016) 264 -- 273.
\newblock \href {https://doi.org/10.1016/j.optcom.2016.01.068}
  {\path{doi:10.1016/j.optcom.2016.01.068}}.

\bibitem{Zhai:16}
D.~Zhai, S.~Chen, S.~Xue, Z.~Yin, Exact recovery of wavefront from
  multishearing interferograms in spatial domain, Appl. Opt. 55~(28) (2016)
  8063--8069.
\newblock \href {https://doi.org/10.1364/AO.55.008063}
  {\path{doi:10.1364/AO.55.008063}}.

\bibitem{Hudgin:77}
R.~H. Hudgin, Wave-front reconstruction for compensated imaging, J. Opt. Soc.
  Am. 67~(3) (1977) 375--378.
\newblock \href {https://doi.org/10.1364/JOSA.67.000375}
  {\path{doi:10.1364/JOSA.67.000375}}.

\bibitem{Freischlad:86}
K.~R. Freischlad, C.~L. Koliopoulos, Modal estimation of a wave front from
  difference measurements using the discrete fourier transform, J. Opt. Soc.
  Am. A 3~(11) (1986) 1852--1861.
\newblock \href {https://doi.org/10.1364/JOSAA.3.001852}
  {\path{doi:10.1364/JOSAA.3.001852}}.

\bibitem{Servin:96}
M.~Servin, D.~Malacara, J.~L. Marroquin, Wave-front recovery from two
  orthogonal sheared interferograms, Appl. Opt. 35~(22) (1996) 4343--4348.
\newblock \href {https://doi.org/10.1364/AO.35.004343}
  {\path{doi:10.1364/AO.35.004343}}.

\bibitem{Guo:12}
Y.~feng Guo, H.~Chen, J.~Xu, J.~Ding, Two-dimensional wavefront reconstruction
  from lateral multi-shear interferograms, Opt. Express 20~(14) (2012)
  15723--15733.
\newblock \href {https://doi.org/10.1364/OE.20.015723}
  {\path{doi:10.1364/OE.20.015723}}.

\bibitem{Guo:14}
Y.~Guo, J.~Xia, J.~Ding, Recovery of wavefront from multi-shear interferograms
  with different tilts, Opt. Express 22~(10) (2014) 11407--11416.
\newblock \href {https://doi.org/10.1364/OE.22.011407}
  {\path{doi:10.1364/OE.22.011407}}.

\bibitem{Elster:99}
C.~Elster, I.~Weing\"{a}rtner, Solution to the shearing problem, Appl. Opt.
  38~(23) (1999) 5024--5031.
\newblock \href {https://doi.org/10.1364/AO.38.005024}
  {\path{doi:10.1364/AO.38.005024}}.

\bibitem{Elster:99-B}
C.~Elster, I.~Weing\"{a}rtner, Exact wave-front reconstruction from two lateral
  shearing interferograms, J. Opt. Soc. Am. A 16~(9) (1999) 2281--2285.
\newblock \href {https://doi.org/10.1364/JOSAA.16.002281}
  {\path{doi:10.1364/JOSAA.16.002281}}.

\bibitem{Elster:00}
C.~Elster, Exact two-dimensional wave-front reconstruction from lateral
  shearing interferograms with large shears, Appl. Opt. 39~(29) (2000)
  5353--5359.
\newblock \href {https://doi.org/10.1364/AO.39.005353}
  {\path{doi:10.1364/AO.39.005353}}.

\bibitem{Dubra:04}
A.~Dubra, C.~Paterson, C.~Dainty, Wave-front reconstruction from shear phase
  maps by use of the discrete fourier transform, Appl. Opt. 43~(5) (2004)
  1108--1113.
\newblock \href {https://doi.org/10.1364/AO.43.001108}
  {\path{doi:10.1364/AO.43.001108}}.

\bibitem{Falldorf:07}
C.~Falldorf, Y.~Heimbach, C.~von Kopylow, W.~J\"{u}ptner, Efficient
  reconstruction of spatially limited phase distributions from their sheared
  representation, Appl. Opt. 46~(22) (2007) 5038--5043.
\newblock \href {https://doi.org/10.1364/AO.46.005038}
  {\path{doi:10.1364/AO.46.005038}}.

\bibitem{Zhai:17}
D.~Zhai, S.~Chen, F.~Shi, High spatial resolution zonal reconstruction with
  modified multishear method in frequency domain, Appl. Opt. 56~(29) (2017)
  8067--8074.
\newblock \href {https://doi.org/10.1364/AO.56.008067}
  {\path{doi:10.1364/AO.56.008067}}.

\bibitem{Zhai:17-high}
D.~Zhai, S.~Chen, F.~Shi, Z.~Yin, Exact multi-shear reconstruction method with
  different tilts in spatial domain, Optics Communications 402 (2017) 453 --
  461.
\newblock \href {https://doi.org/10.1016/j.optcom.2017.06.031}
  {\path{doi:10.1016/j.optcom.2017.06.031}}.

\bibitem{Tomioka:23}
S.~Tomioka, N.~Miyamoto, Y.~Yamauchi, Y.~Matsumoto, S.~Heshmat, Wavefront
  restoration from lateral shearing data using spectral interpolation, Appl.
  Opt. 62~(29) (2023) 7549--7559.
\newblock \href {https://doi.org/10.1364/AO.500453}
  {\path{doi:10.1364/AO.500453}}.

\bibitem{Tian:95}
X.~Tian, M.~Itoh, T.~Yatagai, Simple algorithm for large-grid phase
  reconstruction of lateral-shearing interferometry, Appl. Opt. 34~(31) (1995)
  7213--7220.
\newblock \href {https://doi.org/10.1364/AO.34.007213}
  {\path{doi:10.1364/AO.34.007213}}.

\bibitem{Takeda:82}
M.~Takeda, H.~Ina, S.~Kobayashi, Fourier-transform method of fringe-pattern
  analysis for computer-based topography and interferometry, J. Opt. Soc. Am.
  72~(1) (1982) 156--160.
\newblock \href {https://doi.org/10.1364/JOSA.72.000156}
  {\path{doi:10.1364/JOSA.72.000156}}.

\bibitem{Tomioka:12}
S.~Tomioka, S.~Nishiyama, Phase unwrapping for noisy phase map using localized
  compensator, Appl. Opt. 51~(21) (2012) 4984--4994.
\newblock \href {https://doi.org/10.1364/AO.51.004984}
  {\path{doi:10.1364/AO.51.004984}}.

\bibitem{Samia:22}
S.~Heshmat, S.~Tomioka, S.~Nishiyama, A.~Hirokami, Localized compensator phase
  unwrapping algorithm based on flux conservable solver, Journal of
  Computational Science 62 (2022) 101752.
\newblock \href {https://doi.org/10.1016/j.jocs.2022.101752}
  {\path{doi:10.1016/j.jocs.2022.101752}}.

\end{thebibliography}
\end{document}